\documentclass[a4paper,11pt]{article}
\usepackage{pos}
\usepackage{dsfont}

\makeatletter
\@addtoreset{equation}{section}

\makeatother

\title{Tensor renormalization group study of the two-dimensional lattice U(1) gauge-Higgs model with a topological $\theta$ term under L{\"u}scher's admissibility condition}
\ShortTitle{TRG study of the $2d$ lattice U(1) gauge-Higgs model}

\author*[a,b]{Shinichiro Akiyama}
\author[a]{Yoshinobu Kuramashi}

\affiliation[a]{Center for Computational Sciences, University of Tsukuba,\\
  Tsukuba, Ibaraki 305-8577, Japan}

\affiliation[b]{Graduate School of Science, The University of Tokyo,\\
  Bunkyo-ku, Tokyo 113-0033, Japan}

\emailAdd{akiyama@ccs.tsukuba.ac.jp}
\emailAdd{kuramasi@het.ph.tsukuba.ac.jp}

\notes{
\note{UTHEP-799, UTCCS-P-163}
}

\abstract{
We investigate the two-dimensional lattice U(1) gauge-Higgs model with a topological term, employing L{\"u}scher's admissibility condition.
The standard Monte Carlo simulation for this model is hindered not only by the complex action problem due to the topological term but also by the topological freezing problem originating from the admissibility condition.
Resolving both obstacles simultaneously with the tensor renormalization group approach, we show the advantage of the admissibility condition in dealing with the topological term discretized with the so-called field-theoretical definition.
}

\FullConference{The 41st International Symposium on Lattice Field Theory (LATTICE2024)\\
 28 July - 3 August 2024\\
Liverpool, UK\\}

\begin{document}
\maketitle

\section{Introduction}

In investigating the topological properties of quantum field theories from lattice theories, L{\"u}scher's admissibility condition~\cite{Luscher:1998du}, which gives a formulation of lattice gauge theory that preserves the topological structure, is of crucial importance.
By the admissibility condition, gauge fields are separated into disconnected subspaces corresponding to the different topological sectors.
The admissibility condition reads
\begin{align}
\label{eq:admissibility}
    ||1-P_{\mu\nu}(n)||<\epsilon,~~~
    \forall n, \mu, \nu,
\end{align}
where $\epsilon$ is some fixed positive number and $P_{\mu\nu}(n)$ is a plaquette defined as a product of link variables $U_{\mu}(n)$ such as
$
    P_{\mu\nu}(n)
    =
    U_{\mu}(n)U_{\nu}(n+\hat{\mu})U^{\dag}_{\mu}(n+\hat{\nu})U^{\dag}_{\nu}(n)
$.
L{\"u}scher introduced the following gauge action that automatically meets the admissibility condition in Eq.~\eqref{eq:admissibility},
\begin{align}
\label{eq:luscher_action}
    S_{g} =
    \begin{cases}
        \displaystyle
        \beta\sum_{n,\mu>\nu}
        \frac{1-{\rm Re}P_{\mu\nu}(n)}{1-\| 1-P_{\mu\nu}(n)\|/\epsilon}\;\; 
        & {\rm if}\;\; \| 1-P_{\mu\nu}(n) \| < \epsilon,\\
        \infty & {\rm otherwise},
    \end{cases}
\end{align}
where $\beta$ denotes an inverse gauge coupling.

Although the L{\"u}scher gauge action in Eq.~\eqref{eq:luscher_action} should be superior to the standard Wilson gauge action, particularly in focusing on the topological phenomena from the lattice gauge theories, the Monte Carlo simulation based on the L{\"u}scher gauge action encounters severe difficulty unfortunately.
As reported in Ref.~\cite{Fukaya:2003ph}, the topological change is substantially suppressed under the Monte Carlo updates.
This issue is usually referred to as the topological freezing problem.
This is why the Monte Carlo simulation for the L{\"u}scher gauge action is performed in the fixed topological sectors~\cite{Fukaya:2004kp,Bietenholz:2005rd,Bietenholz:2016ymo}.

We demonstrate that this obstacle is resolved by the tensor renormalization group (TRG) which approximately computes the path integral expressed as a tensor network.
Since the TRG evaluates the path integral itself, the numerical results obtained by the TRG automatically include all the contributions from different topological sectors.
Additionally, there is no difficulty in imposing the periodic boundary condition, suitable for dealing with topological terms, in the TRG calculations.
In this study, we investigate the two-dimensional ($2d$) lattice U(1) gauge-Higgs model defined by the L{\"u}scher gauge action with a topological $\theta$ term, particularly the phase transition at $\theta=\pi$ with a sufficiently large mass parameter.
In addition to the topological freezing problem, the standard Monte Carlo simulation suffers from a complex action problem when the model has a topological $\theta$ term.
However, since the TRG is free from the complex action problem associated with a topological $\theta$ term as well~\cite{Shimizu:2014fsa,Kawauchi:2016xng,Kawauchi:2017dnj,Kuramashi:2019cgs,Butt:2019uul,Hirasawa:2021qvh,Nakayama:2021iyp,Kanno:2024elz}, we can investigate the phase transition at $\theta=\pi$ without suffering from these two numerical obstacles.

\section{The model and its tensor network representation}

The action of the U(1) gauge-Higgs model with a topological $\theta$ term is given by
\begin{align}
\label{eq:action}
    S=
    S_{g}+S_{h}+S_{\theta}
    ,
\end{align}
where $S_{g}$ is the L{\"u}scher gauge action in Eq.~\eqref{eq:luscher_action} with the U(1)-valued link variable $U_{\mu}(n)$, and $S_{h}$ defines the Higgs sector as
\begin{align}
    S_{h}
    &=
    -
    \sum_{n,\mu}
    \left[
        \phi^{*}(n)
        U_{\mu}(n)
        \phi(n+\hat{\mu})
        +
        \phi^{*}(n+\hat{\mu})
        U^{*}_{\mu}(n)
        \phi(n)
    \right]
    +M\sum_{n}\left|\phi(n)\right|^{2}
    +\lambda\sum_{n}\left|\phi(n)\right|^{4}
    .
\end{align}
The Higgs field is denoted by the single-component complex-valued field $\phi(n)$ and the lattice mass parameter by $M$. The quartic coupling is given by $\lambda$.
The topological $\theta$ term is introduced by the third term in Eq.~\eqref{eq:action}.
One way to define a topological $\theta$ term on a lattice is 
\begin{align}
\label{eq:theta_log}
    S_{\theta}
    =
    \frac{{\rm i}\theta}{2\pi}
    \sum_{n}\ln P_{12}(n)
    ,
\end{align}
which grantees the $2\pi$ periodicity with respect to $\theta$ not only in the continuum limit but also in the lattice model.
Alternatively, 
\begin{align}
\label{eq:theta_sin}
    S_{\theta}
    =
    \frac{{\rm i}\theta}{2\pi}
    \sum_{n}{\rm Im} P_{12}(n)
    ,
\end{align}
is also a widely used action of a topological $\theta$ term on a lattice, known as the field-theoretical definition.
Although this definition grants the $2\pi$ periodicity only in the continuum limit, a topological term on a lattice in four dimensions can be defined similarly to Eq.~\eqref{eq:theta_sin}.

We parametrize the link variable via $U_{\mu}(n)={\rm e}^{{\rm i}\vartheta_{\mu}(n)}$ with $\vartheta_{\mu}(n)\in[-\pi,\pi]$ and employ the polar coordinate for the Higgs field as $\phi(n)=r(n){\rm e}^{{\rm i}\varphi(n)}$.
Thanks to the invariance of the Haar measure, the angular field $\varphi(n)$ can be eliminated from the path integral, which reads
\begin{align}
\label{eq:path_integral}
    Z=
    \prod_{n,\mu}\int^{\pi}_{-\pi}\frac{{\rm d}\vartheta_{\mu}(n)}{2\pi}
    \prod_{n}\int^{\infty}_{0}r(n){\rm d}r(n)
    \exp\left[-\beta S_{g}-S'_{h}-S_{\theta}\right]
    ,
\end{align}
where
\begin{align}
    S'_{h}
    &=
    -
    \sum_{n,\mu}
    2r(n)r(n+\hat{\mu})
    \cos\vartheta_{\mu}(n)
    +\sum_{n}\left[Mr(n)^{2}+\lambda r(n)^{4}\right]
    .
\end{align}

We now represent the path integral in Eq.~\eqref{eq:path_integral} as a tensor network.
We assume the model is defined on a square lattice with periodic boundary conditions for both two directions.
Since Eq.~\eqref{eq:path_integral} includes the integrations over the continuous variables, they have to be replaced by summations, otherwise Eq.~\eqref{eq:path_integral} cannot be expressed as a tensor contraction.
In this study, we employ the Gauss-Legendre quadrature for the integrations over $\vartheta_{\mu}(n)$ and the Gauss-Laguerre quadrature for those over $r(n)$, where the roots of Legendre and Laguerre polynomials are used to determine the sampling points, respectively.
Denoting the number of sampling points for the link variables as $K_{g}$ and that for the Higgs fields as $K_{h}$, we can approximately represent the original path integral as a tensor network via
\begin{align}
\label{eq:tn_rep}
    Z\simeq Z(K_{g},K_{h})
    =
    {\rm tTr}\left[
        \prod_{n}T_{n}
    \right]
    ,
\end{align}
where $T_{n}$ is a fundamental tensor defined on each lattice site $n$.
We emphasize that the L{\"u}scher gauge action in Eq.~\eqref{eq:luscher_action} is straightforwardly implemented in the fundamental tensor $T_{n}$ when we apply the Gauss quadrature as a discretization scheme.
For the details on this tensor network representation, see Ref.~\cite{Akiyama:2024qer}.
Although the original path integral is restored in the limits of $K_{g}\to\infty$ and $K_{h}\to\infty$, the Gauss quadratures provide us with sufficiently accurate results with finite $K_{g}$ and $K_{h}$, as we will see below.

\section{Numerical results}

We use the bond-weighted TRG (BTRG)~\cite{PhysRevB.105.L060402}, which improves the original TRG~\cite{Levin:2006jai} without increasing the computational cost, to perform the tensor contractions in Eq.~\eqref{eq:tn_rep}.
BTRG enables us to approximately carry out the contractions among $2^{p}$ fundamental tensors, which corresponds to the lattice size $V$ via $V=2^{p}$, just within $p$ times of coarse-graining transformation constructed by the truncated singular value decomposition, whose accuracy is controlled by the bond dimension $D_{\rm BTRG}$.
This is another algorithmic parameter in addition to $K_{g}$ and $K_{h}$ in the Gauss quadratures.
The quartic coupling is fixed as $\lambda=0.5$ throughout this study.
For the algorithmic parameters, we always set $(K_{g},K_{h},D_{\rm BTRG})=(20,20,160)$.

\begin{figure}[htbp]
    \centering
    \begin{minipage}[t]{0.49\hsize}
        \includegraphics[width=1.0\hsize]{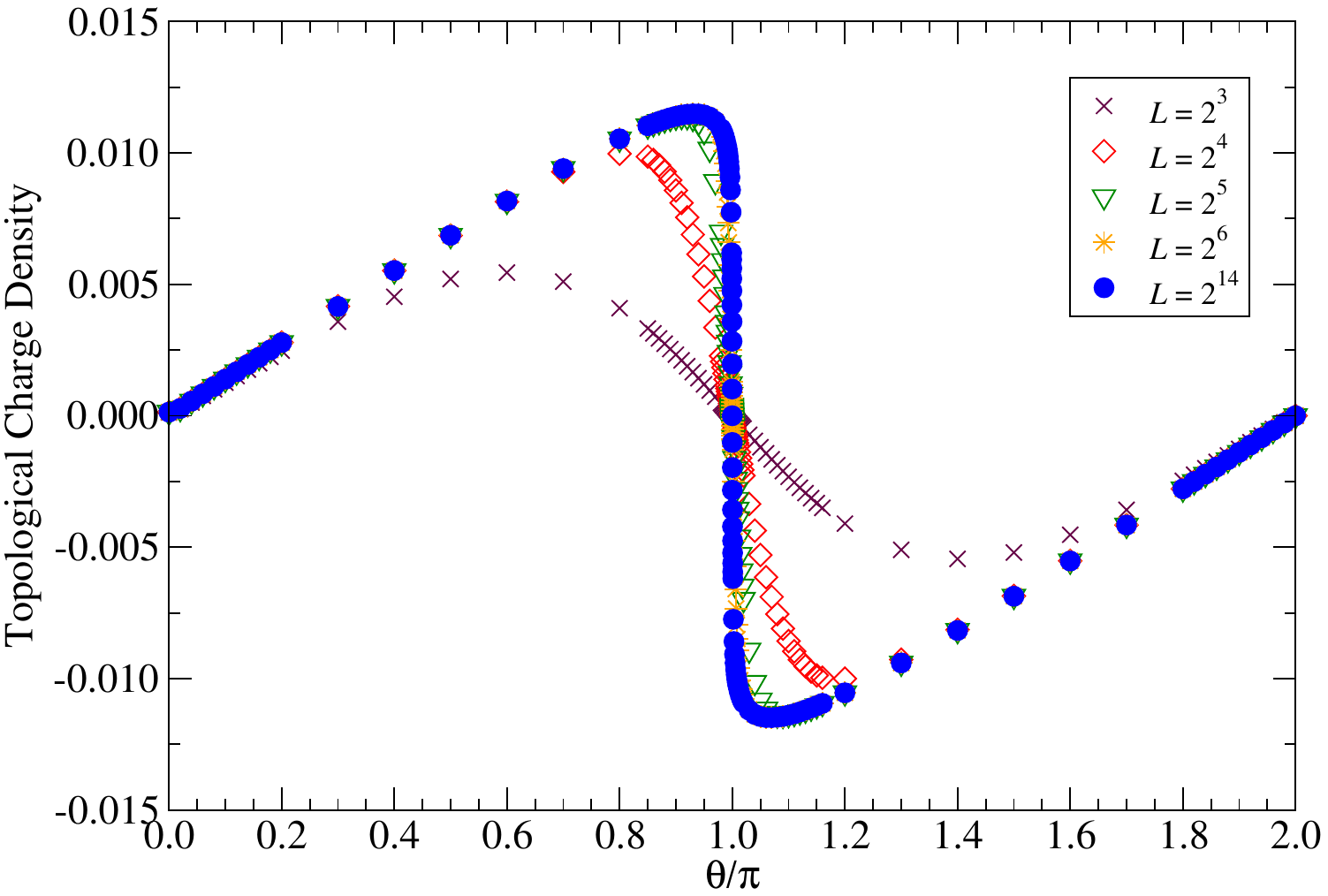}
    \end{minipage} 
    \begin{minipage}[t]{0.49\hsize}
        \includegraphics[width=1.0\hsize]{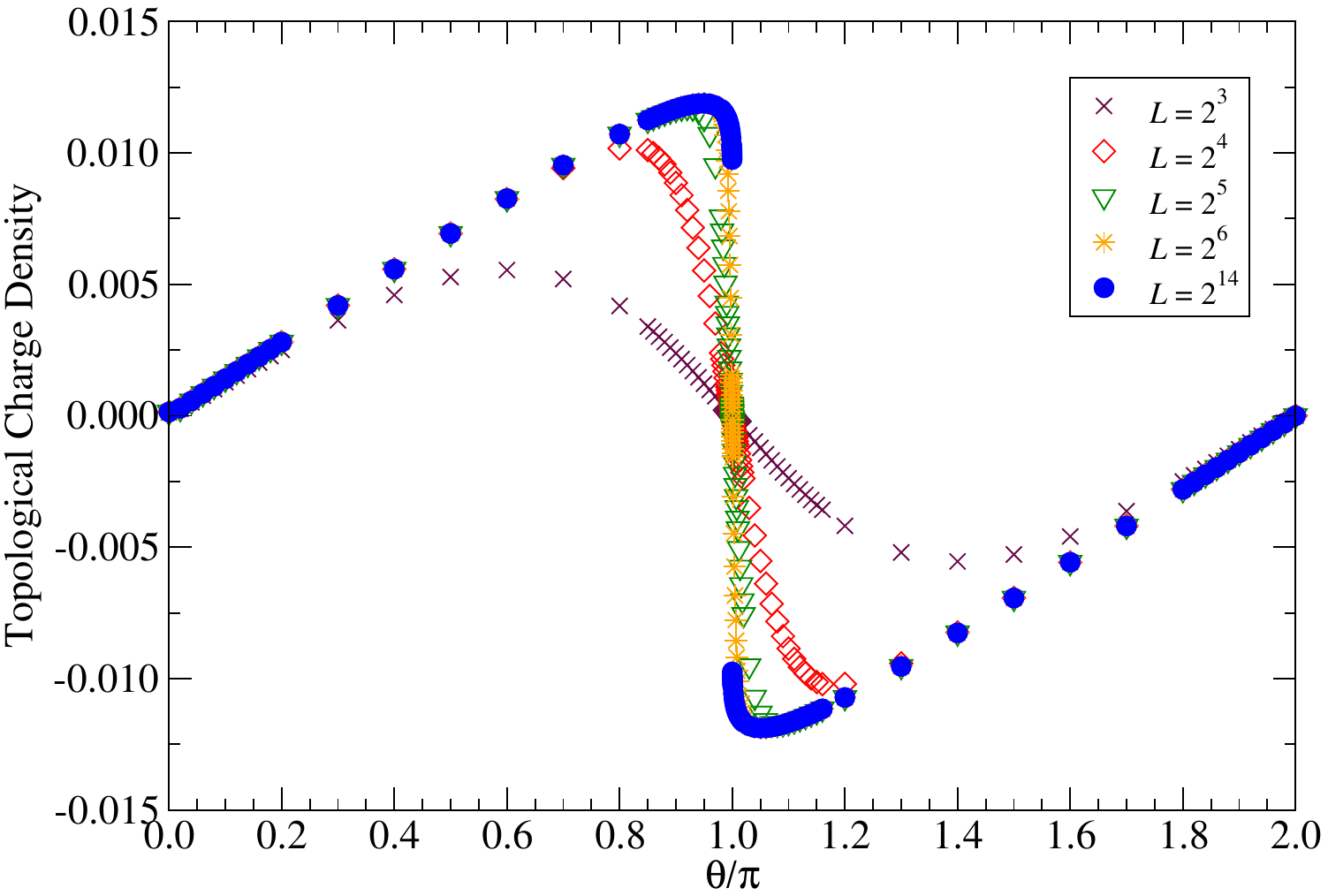}
    \end{minipage} 
    \caption{
        Topological charge density as a function of $\theta/\pi$ at $\beta=3$, $\epsilon=1$, and $\lambda=0.5$ with $M=2.99$ (left), $M=3.00$ (right) at various lattice volumes. 
    }
    \label{fig:top_charge_log}
\end{figure}

\begin{figure}[htbp]
    \begin{tabular}{p{0.45\textwidth}p{0.03\textwidth}p{0.45\textwidth}}
        \centering
        \includegraphics[width=1\hsize]{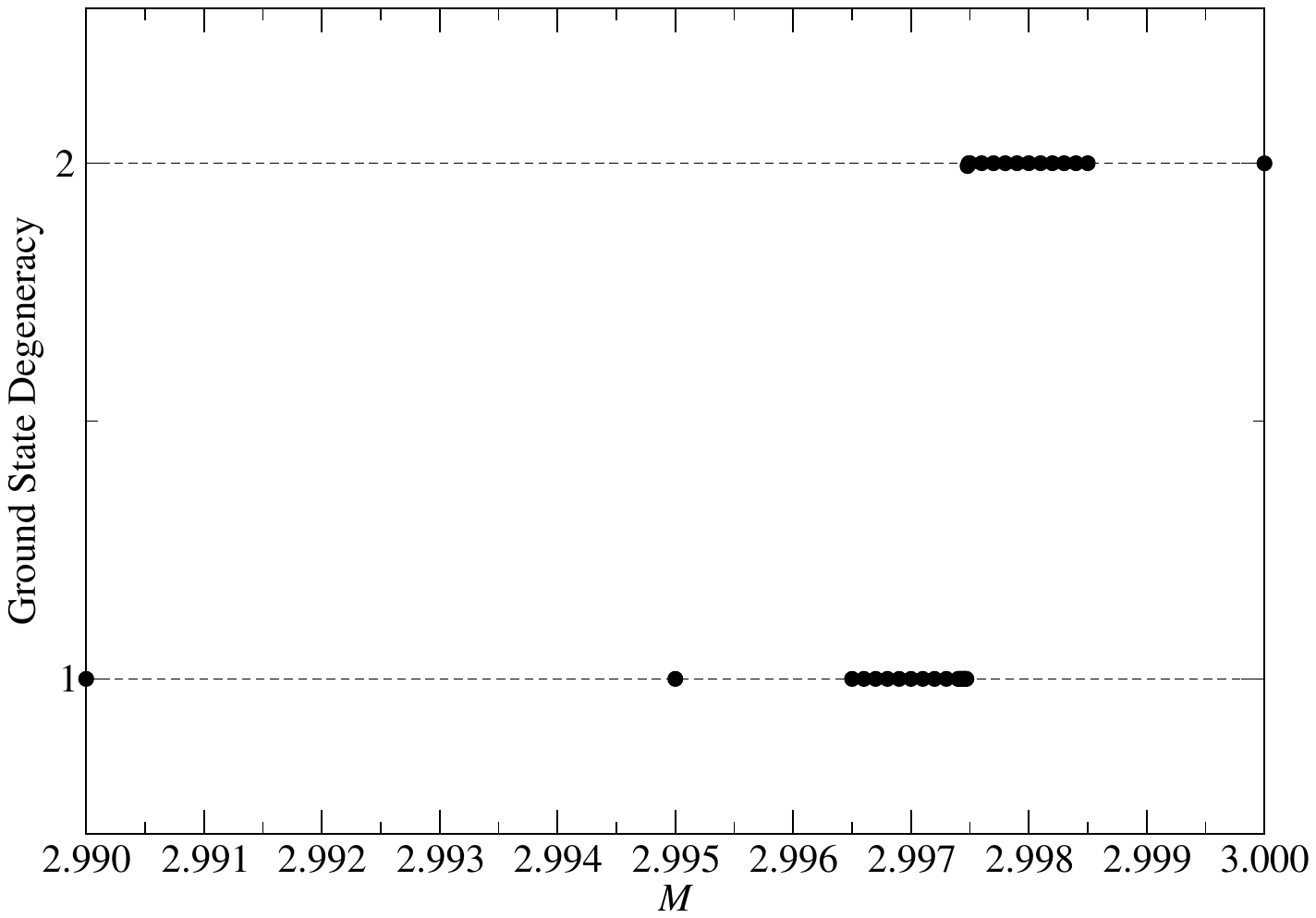}
        \caption{
            Ground state degeneracy at $\theta=\pi$ as a function of $M$ at $V=2^{40}$.
        }
        \label{fig:deg_x}
    &&
        \centering
        \includegraphics[width=1\hsize]{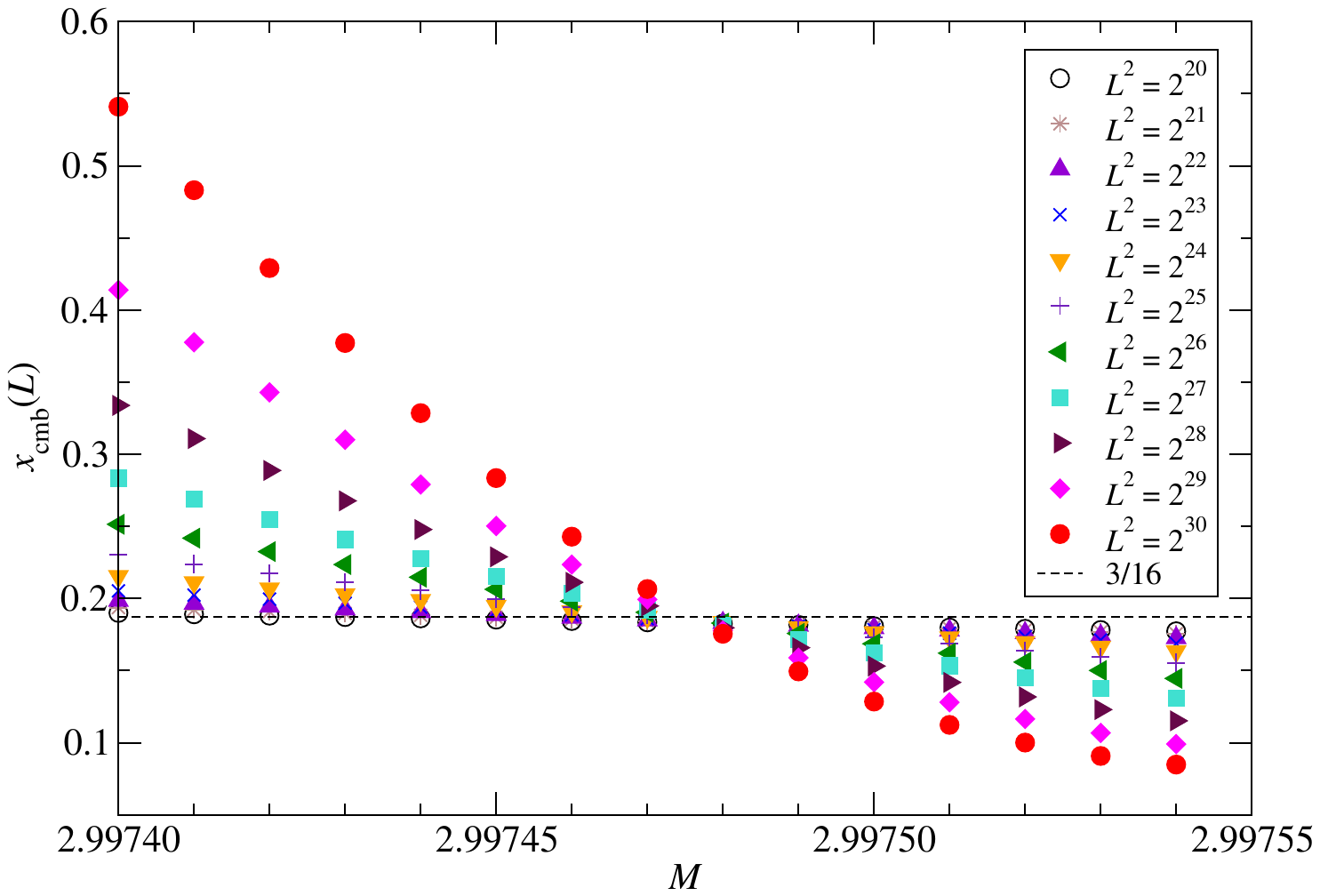}
        \caption{
            The volume dependence of the combined scaling dimension $x_{\rm cmb}(L)$.
            The dashed line denotes $x_{\rm cmb}=3/16$.
        }
        \label{fig:scaling_cmb}
    \end{tabular}
\end{figure}

Firstly, we investigate the model with the topological $\theta$ term defined by Eq.~\eqref{eq:theta_log}.
Fig.~\ref{fig:top_charge_log} shows the topological charge density, defined by
\begin{align}
    \frac{\langle Q\rangle}{V}
    =
    -\frac{1}{V}
    \frac{\partial \ln Z}{\partial\theta}
    ,
\end{align}
at $\beta=3$ and $\epsilon=1$.
The resulting topological charge density varies smoothly against $\theta$ with $M=2.99$.
On the other hand, it is discontinuous at $\theta=\pi$ when $M=3.00$.
This is consistent with the expected phase structure in Ref.~\cite{Komargodski:2017dmc}; the model exhibits the first-order transition at $\theta=\pi$ with the spontaneous breaking of the $\mathds{Z}_{2}$ charge conjugation symmetry when the Higgs mass parameter $M$ is sufficiently large.
This first-order transition at $\theta=\pi$ terminates at some critical mass $M_{\rm c}$, whose critical behavior is expected to be in the $2d$ Ising universality class.
To locate the critical endpoint, we compute the ground state degeneracy according to Ref.~\cite{PhysRevB.80.155131}.
Fig.~\ref{fig:deg_x} shows the ground state degeneracy as a function of $M$, which gives the bound for $M_{\rm c}$ as $2.99747\le M_{\rm c}\le2.99748$.
We also employ the recently proposed tensor-network-based level spectroscopy~\cite{PhysRevB.104.165132,PhysRevB.108.024413} to precisely locate the critical endpoint, identifying its universality class.
Let $\lambda_{n}~(n=0,1,2,\cdots)$ be an $n$-th eigenvalue of the transfer matrix.
Assuming that these eigenvalues are in descending order, the scaling dimension $x_{n}(L)$ with the finite system size $L=V^{1/2}$ is given by
\begin{align}
    x_{n}(L)=\frac{1}{2\pi}\ln\frac{\lambda_{0}(L)}{\lambda_{n}(L)}.
\end{align}
Ref.~\cite{PhysRevB.108.024413} utilizes a linear combination of two scaling dimensions such as $x_{\rm cmb}(L)=x_{1}(L)+x_{2}(L)/16$ to remove the effect from the dominant irrelevant perturbation with the scaling dimension 4.
Fig.~\ref{fig:scaling_cmb} shows the combined scaling dimension $x_{\rm cmb}(L)$ at various system sizes as a function of $M$.
The almost volume-independent behavior with $x_{\rm cmb}=3/16$ is observed at $M\sim2.99748$, which is a clear signal of the $2d$ Ising universality.
Following the strategy of the tensor-network-based level spectroscopy~\cite{PhysRevB.104.165132,PhysRevB.108.024413}, the critical endpoint is finally located as $M_{\rm c}=2.9974765(14)$.
\footnote{For the algorithmic parameter dependence of the resulting critical endpoint, see Ref.~\cite{Akiyama:2024qer}.}
This estimation is comparable with $M_{\rm c}=2.989(2)$ obtained in Ref.~\cite{Gattringer:2018dlw} by the Monte Carlo simulation based on the dual representation, where the Villain form defines the Boltzmann weight for the gauge fields.
We also investigate the finite-size correction of the free energy density $\ln Z/V$ to obtain the central charge $c$.
Using the data at $M=2.99748$ with $L\in[2^{10},2^{15}]$, the non-vanishing central charge is obtained as $c=0.50(7)$, which is another evidence of the emergence of the $2d$ Ising universality class.

\begin{figure}[htbp]
    \centering
    \begin{minipage}[t]{0.49\hsize}
        \includegraphics[width=1.0\hsize]{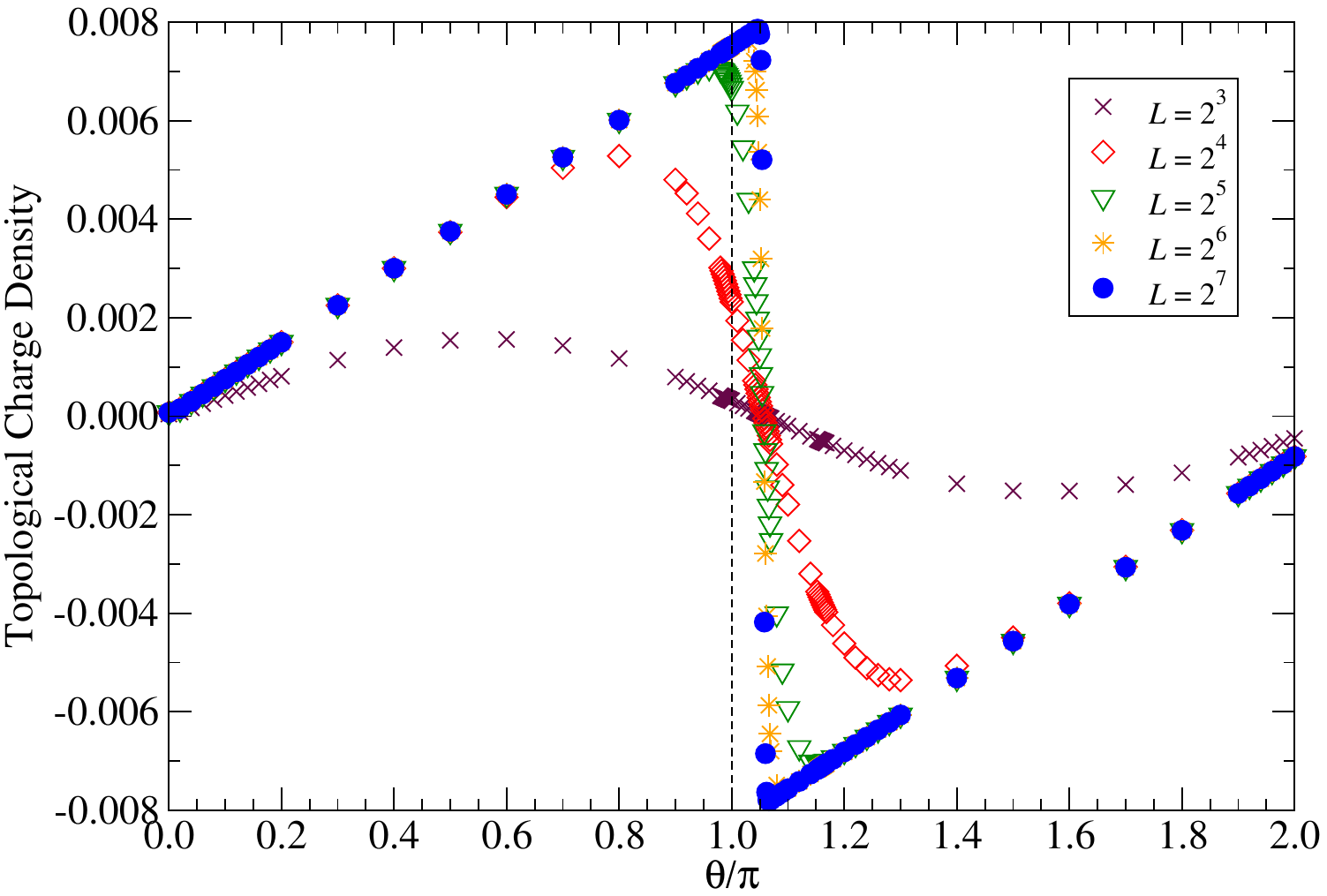}
    \end{minipage} 
    \begin{minipage}[t]{0.49\hsize}
        \includegraphics[width=1.0\hsize]{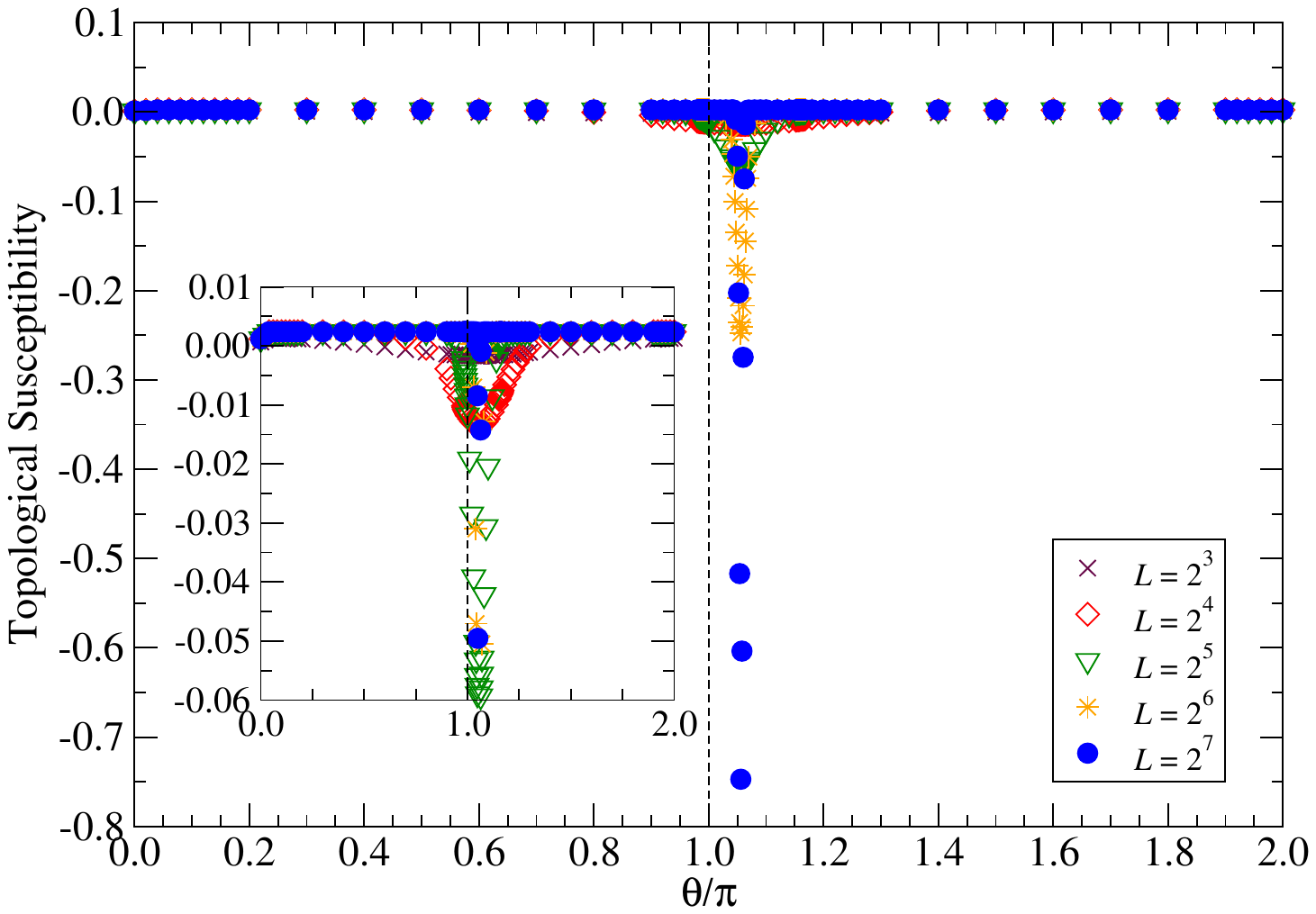}
    \end{minipage} 
    \caption{
        Topological charge density (left) and topological susceptibility (right) as a function of $\theta/\pi$ at $\beta=10$, $M=4$ with the standard Wilson gauge action at various lattice volumes. 
        Dashed vertical lines denote $\theta=\pi$.
        The inset graph is provided for the susceptibility in the smaller volumes.
    }
    \label{fig:top_quantities_sin_wilson}
\end{figure}

Next, we consider the same model but with the topological $\theta$ term defined by Eq.~\eqref{eq:theta_sin}.
We begin with the computation using the standard Wilson gauge action instead of the L{\"u}scher gauge action because the numerical results can be directly compared with a previous dual simulation provided in Ref.~\cite{Gattringer:2015baa}.
Fig.~\ref{fig:top_quantities_sin_wilson} shows the topological charge density and its susceptibility, defined by
\begin{align}
    \chi_{Q}
    =
    -\frac{1}{V}
    \frac{\partial^{2} \ln Z}{\partial\theta^{2}}
    ,
\end{align}
at $\beta=10$ and $M=4$.
These results shown in Fig.~\ref{fig:top_quantities_sin_wilson} seem quantitatively consistent with Figure 11 in Ref.~\cite{Gattringer:2015baa}.
Since a clear discontinuity is observed in the topological charge density with $L=2^{7}$, it seems that the first-order transition takes place when $M=4$ but the transition point deviates from $\theta=\pi$ due to the lattice artifact.

\begin{figure}[htbp]
    \centering
    \begin{minipage}[t]{0.49\hsize}
        \includegraphics[width=1.0\hsize]{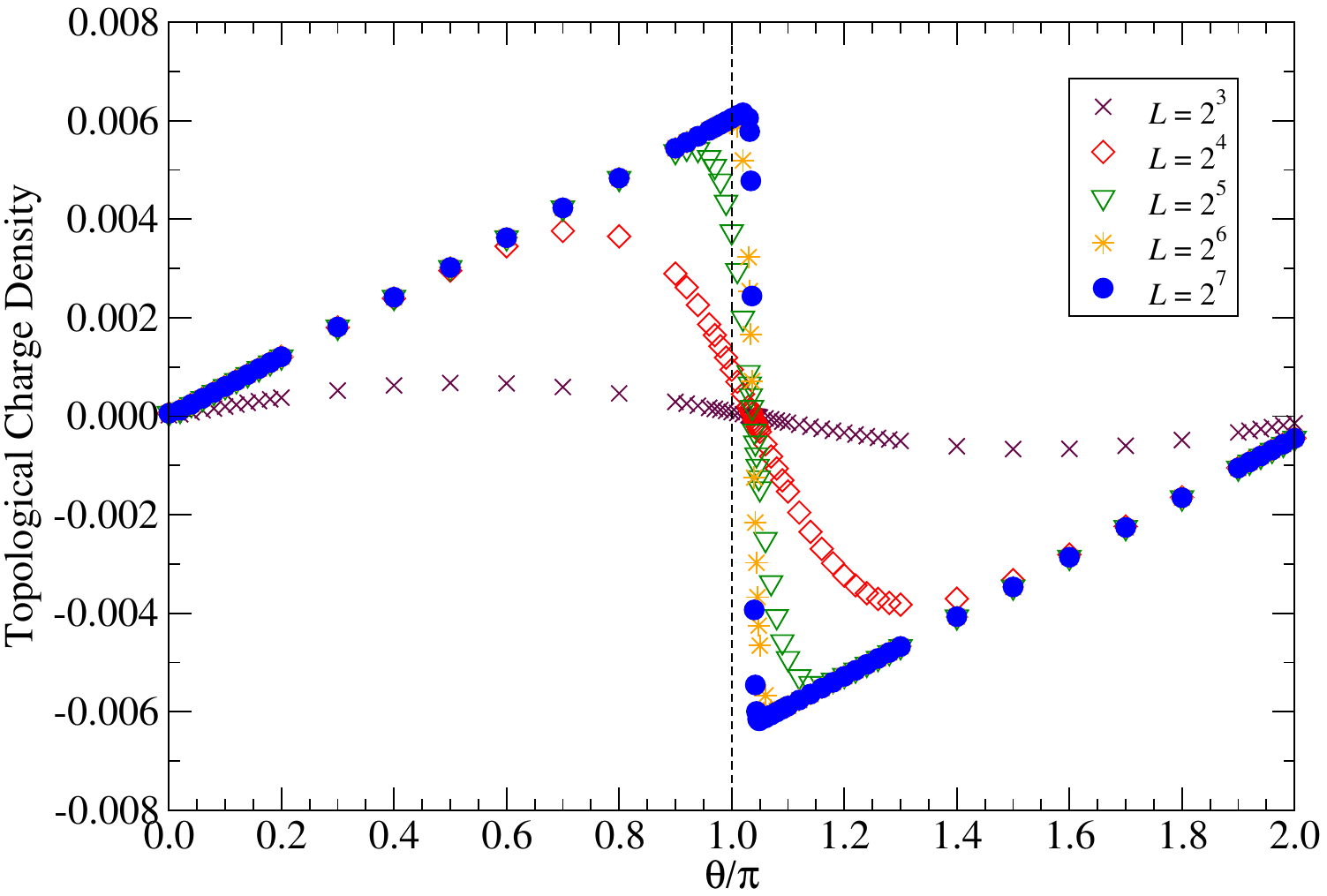}
    \end{minipage} 
    \begin{minipage}[t]{0.49\hsize}
        \includegraphics[width=1.0\hsize]{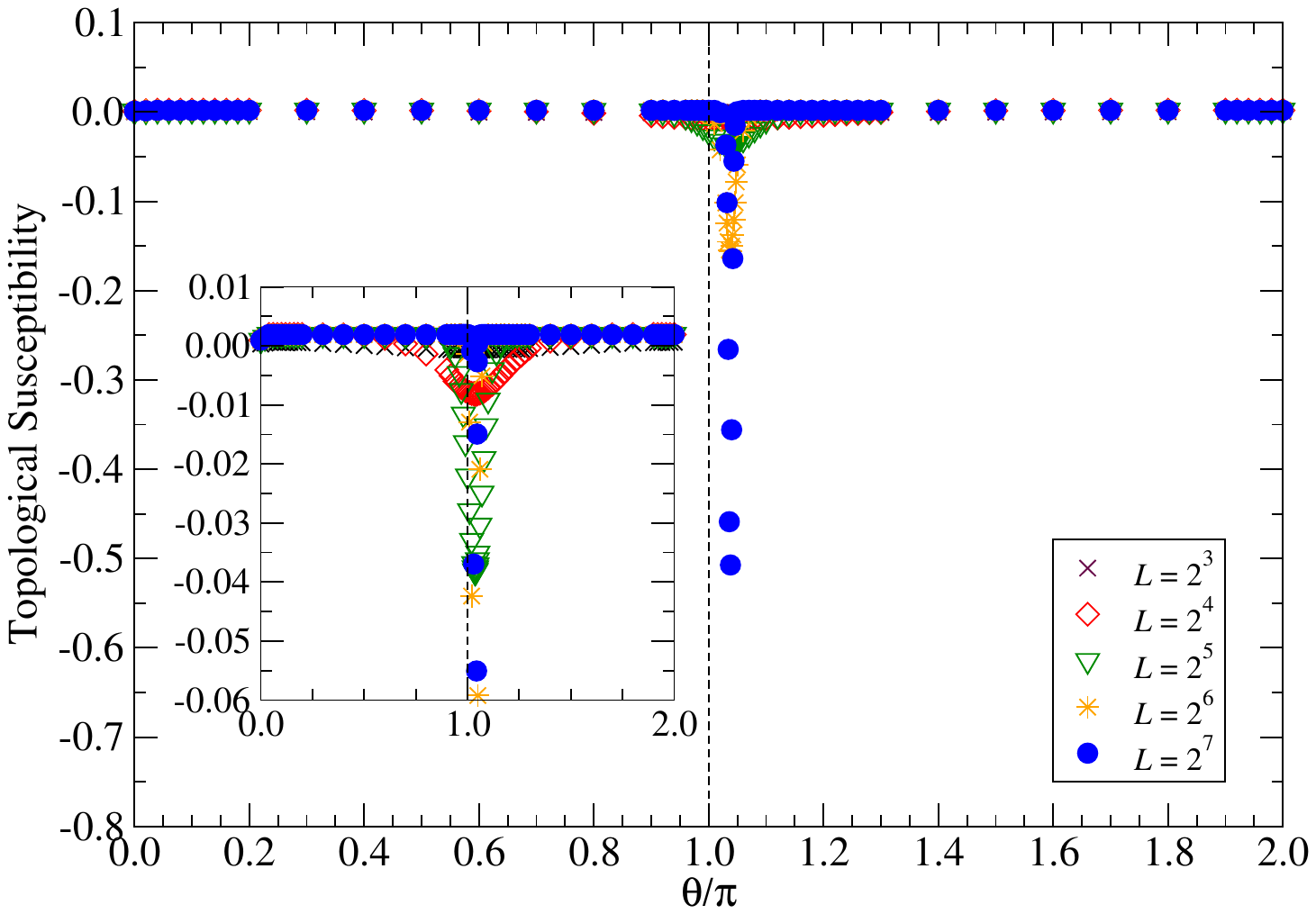}
    \end{minipage} 
    \caption{
        Topological charge density (left) and topological susceptibility (right) as a function of $\theta/\pi$ at $\beta=10$, $M=4$ with the L{\"u}scher gauge action ($\epsilon=1$) at various lattice volumes. 
        Dashed vertical lines denote $\theta=\pi$.
        The graph scale is the same as Fig.~\ref{fig:top_quantities_sin_wilson}.
    }
    \label{fig:top_quantities_sin_epsilon1}
\end{figure} 

\begin{figure}[htbp]
    \centering
    \begin{minipage}[t]{0.49\hsize}
        \includegraphics[width=1.0\hsize]{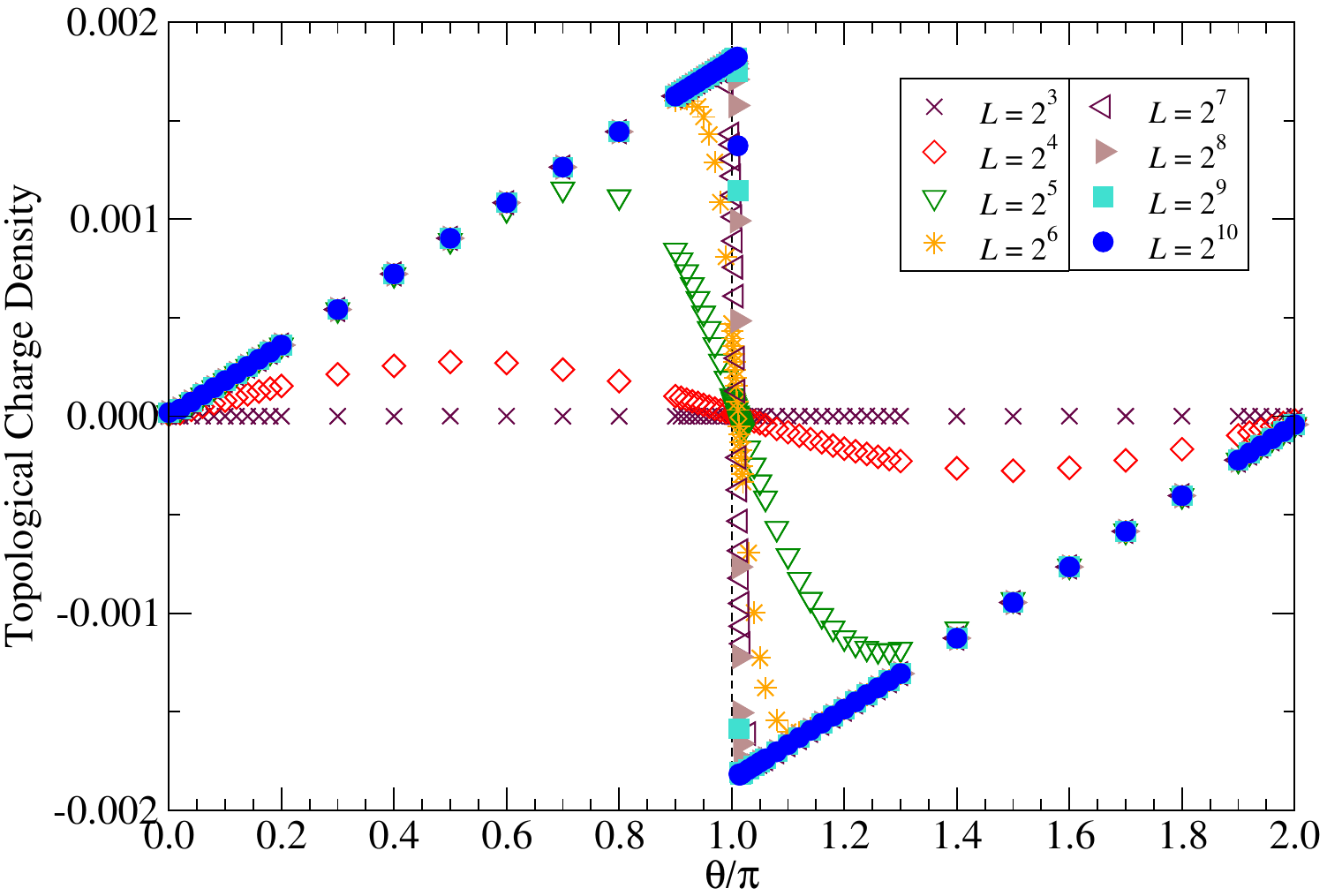}
    \end{minipage} 
    \begin{minipage}[t]{0.49\hsize}
        \includegraphics[width=1.0\hsize]{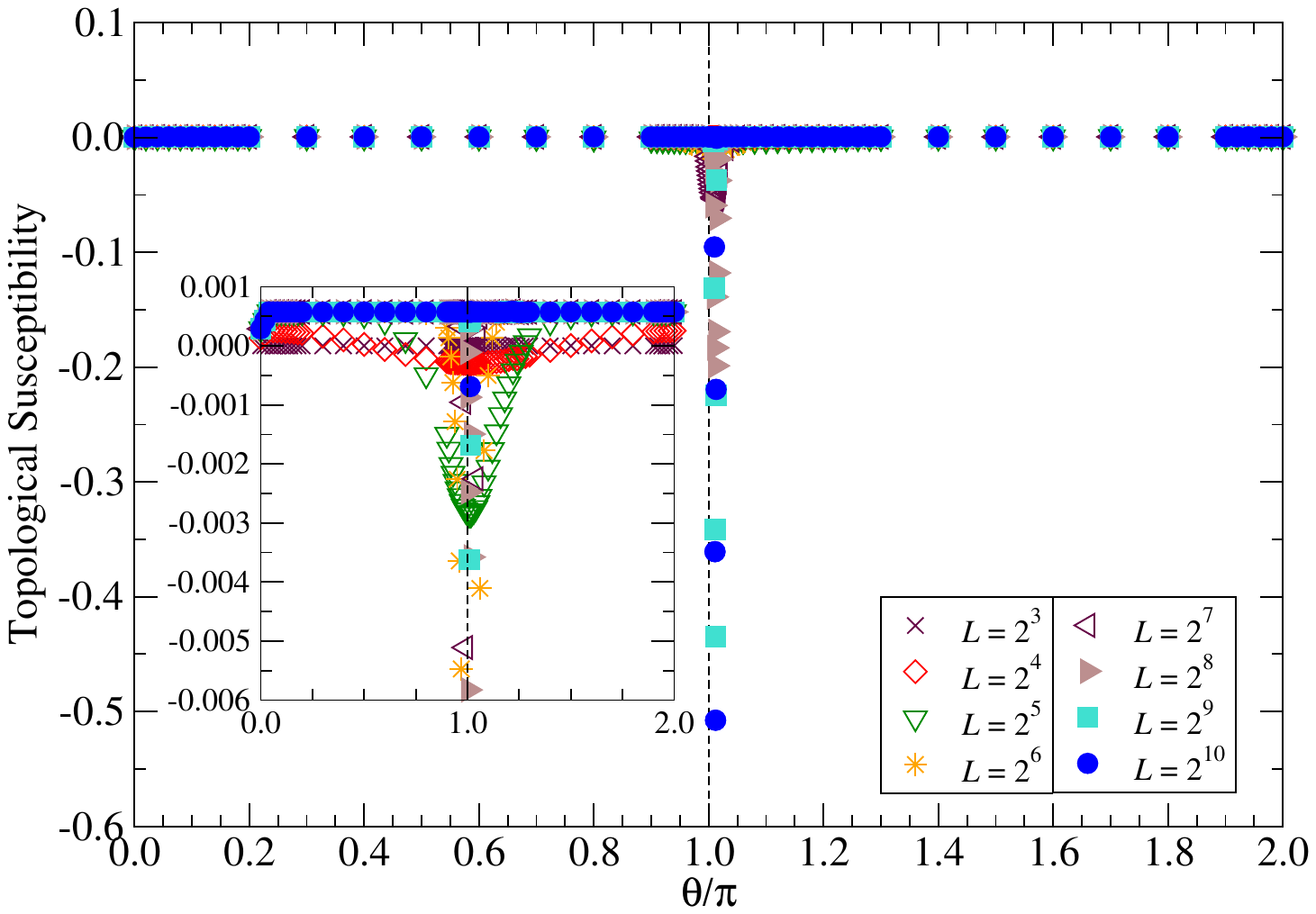}
    \end{minipage} 
    \caption{
        Topological charge density (left) and topological susceptibility (right) as a function of $\theta/\pi$ at $\beta=10$, $M=4$ with the L{\"u}scher gauge action ($\epsilon=0.1$) at various lattice volumes. 
        Dashed vertical lines denote $\theta=\pi$.
        The inset graph is provided for the susceptibility in the smaller volumes.
    }
    \label{fig:top_quantities_sin_epsilon01}
\end{figure}

\begin{figure}[htbp]
    \centering
    \includegraphics[width=0.8\hsize]{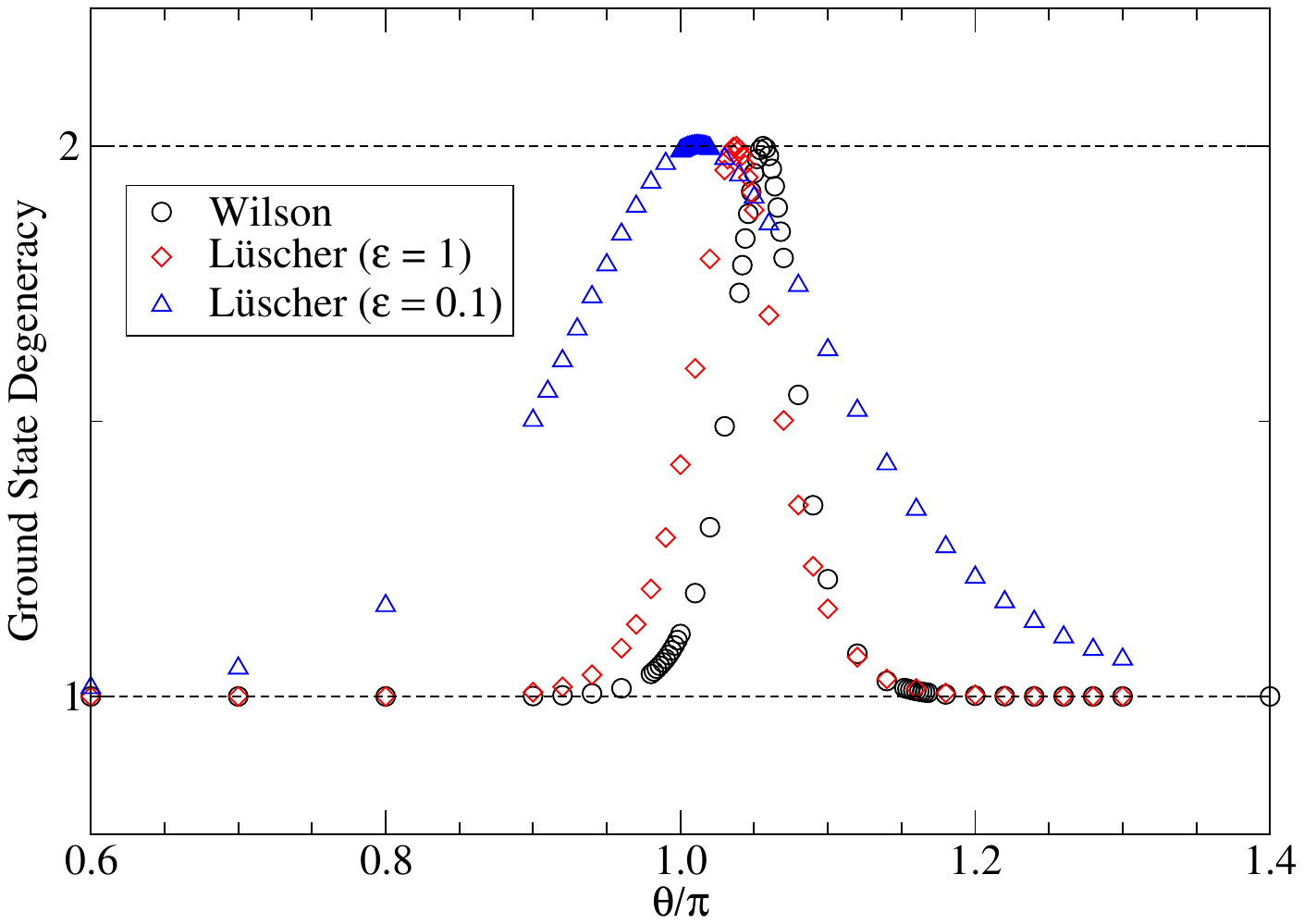}
    \caption{
        Ground state degeneracy as a function of $\theta/\pi$ at $\beta=10$, $M=4$ at $L=2^{5}$. 
        Each symbol denotes the result obtained by the Wilson gauge action, L{\"u}scher gauge action with $\epsilon=1$, and $\epsilon=0.1$.
    }
    \label{fig:deg_x_sin}
\end{figure}

Let us now move on to the case with the L{\"u}scher gauge action.
The resulting topological charge density and its susceptibility are shown in
Fig.~\ref{fig:top_quantities_sin_epsilon1}, where we set $\epsilon=1$.
The results obtained by the L{\"u}scher gauge action with $\epsilon=1$ are qualitatively consistent with those by the Wilson gauge action.
Compared with Fig.~\ref{fig:top_quantities_sin_wilson}, however, the transition point is now closer to $\theta=\pi$.
This is a clear advantage of the L{\"u}scher gauge action particularly in dealing with the topological $\theta$ term according to the field-theoretical definition.
We further examine how these topological quantities are modified by setting $\epsilon$ even smaller.
In Fig.~\ref{fig:top_quantities_sin_epsilon01}, we have set $\epsilon=0.1$.
It can be seen that the resulting transition point is almost at $\theta=\pi$.
Although the finite-volume effect seems to be enhanced with smaller $\epsilon$, this does not provide any extra difficulty in the TRG computations because the computational cost of the TRG scales logarithmically with respect to the lattice volume.
With $L\ge2^{9}$, the discontinuity in the topological charge density emerges as well as in the previous cases, indicating the first-order transition.
Finally, we investigate the ground state degeneracy near $\theta=\pi$.
In Fig.~\ref{fig:deg_x_sin}, we show the degeneracy computed on a finite lattice with $L=2^{5}$ as a function of $\theta$.
We note that the two-fold degeneracy should be observed only at the transition point in the thermodynamic limit. Noninteger values are just due to the finite-volume effect.
The two-fold degeneracy observed near $\theta=\pi$ is the signal of spontaneous $\mathds{Z}_{2}$ symmetry breaking. 
Therefore, the current TRG computations based on the field-theoretical definition work as well as in the previous calculations with the logarithmic definition in Eq.~\eqref{eq:theta_log}.
Fig.~\ref{fig:deg_x_sin} again demonstrates that the L{\"u}scher gauge action with the smaller $\epsilon$ results in the transition point almost at $\theta=\pi$.

\section{Summary and outlook}

We have shown that the TRG is a promising approach in dealing with the lattice gauge theory under L{\"u}scher's admissibility condition.
Specifically, both the topological freezing problem and complex action problems are simultaneously resolved by the TRG approach.
The phase transition at $\theta=\pi$ in the $2d$ U(1) gauge-Higgs model has been investigated by employing two types of definitions for the topological term on a lattice.
With the logarithmic definition, the lattice model exhibits the phase transition exactly at $\theta=\pi$ when $M\ge M_{\rm c}$.
Using the tensor-network level spectroscopy, we have confirmed that the critical behavior at $M=M_{\rm c}$ is in the $2d$ Ising universality class.
With the field-theoretical definition, on the other hand, the L{\"u}scher gauge action has shown a clear advantage over the standard Wilson gauge action; the transition point is located almost at $\theta=\pi$ even at finite $\beta$ by setting sufficiently small $\epsilon$.
The search for the critical endpoint in the $2d$ U(1) gauge-Higgs model with the topological $\theta$ term based on the field-theoretical definition will be reported elsewhere.

\begin{acknowledgments}
Numerical calculation for the present work was carried out with SQUID at the Cybermedia Center, Osaka University (Project ID: hp240012).
We also used the supercomputer Fugaku provided by RIKEN through the HPCI System Research Project (Project ID: hp230247) and computational resources of Wisteria/BDEC-01, Cygnus, and Pegasus under the Multidisciplinary Cooperative Research Program of Center for Computational Sciences, University of Tsukuba. 
This work is supported by Grants-in-Aid for Scientific Research from the Ministry of Education, Culture, Sports, Science and Technology (MEXT) (Nos. 24H00214, 24H00940). 
SA acknowledges the support from the Endowed Project for Quantum Software Research and Education, the University of Tokyo (\url{https://qsw.phys.s.u-tokyo.ac.jp/}), JSPS KAKENHI Grant Number JP23K13096, the Center of Innovations for Sustainable Quantum AI (JST Grant Number JPMJPF2221), and the Top Runners in Strategy of Transborder Advanced Researches (TRiSTAR) program conducted as the Strategic Professional Development Program for Young Researchers by the MEXT.
\end{acknowledgments}


\bibliographystyle{JHEP}
\bibliography{bib/algorithm,bib/continuous,bib/discrete,bib/formulation,bib/gauge,bib/grassmann,bib/gravity,bib/ref,bib/qsw}

\end{document}